\documentclass[runningheads]{llncs}

\input{psfig.sty}

\begin{document}

\pagestyle{headings}

\mainmatter

\title{Corrections of the NIST Statistical Test Suite for Randomness}

\titlerunning{Corrections of the NIST Statistical Test Suite for Randomness}

\author{Song-Ju Kim
\and Ken Umeno \and
Akio Hasegawa}

\authorrunning{Song-Ju Kim et al.}

\institute{Chaos-based Cipher Chip Project, Presidential Research Fund,\\
               Communications Research Laboratory,
               Incorporated Administrative Agency\\ 
               4-2-1, Nukui-kitamachi, Koganei-shi, Tokyo 184-8795,
               Japan\\
\email{\{songju, umeno, ahase\}@crl.go.jp}\\
}

\maketitle

\begin{abstract}
It is well known that the NIST statistical test suite was used for the
 evaluation of AES candidate algorithms.
We have found that the test setting of Discrete Fourier Transform test and Lempel-Ziv
 test of this test suite are wrong.
We give four corrections of mistakes in the test settings.
This suggests that re-evaluation of the test results should be needed.\\ 

\keywordname{ Pseudo-Random Bit Generator, Statistical Test, Discrete
 Fourier Transform, Lempel-Ziv Compression Algorithm, Cellular Automata}
\end{abstract}

\section{Introduction}

Random and pseudorandom bit generators (RBGs, PRBGs) are used for many purposes
 including cryptographic, modeling, and simulation applications.
For cryptographic purpose, they are required in the construction of 
encryption keys, other cryptographic parameters, and so on.
One of the criteria used to evaluate the Advanced Encryption Standard
 (AES) candidate algorithms was their demonstrated suitability as PRBGs.
That is, the evaluation of their outputs utilizing statistical tests
 should not provide any means by which to computationally distinguish
 them from truly random sources~\cite{ref:AES1,ref:AES2,ref:AES3}.

Cryptographically secure pseudorandom bit generator is defined as a
PRBG that passes the next-bit test~\cite{ref:handbook}.
A PRBG is said to pass the next-bit test if there is no polynomial-time
algorithm which, on input of the first $l$ bits of an output sequence
$s$, can predict the ($l$+1)st bit of $s$ with probability significantly
greater than $\frac{1}{2}$.
It is known that a PRBG passes the next-bit test if and only if it
passes all polynomial-time statistical tests.
Although a few PRBGs such as RSA, BBS are known as cryptographically
secure PRBGs under the assumption that RSA problem and integer
factorization are intractable, it is difficult to prove that some PRBG
is cryptographically secure in general.   
Practically, we only subject a sample output sequence of the PRBG to 
various statistical tests, and evaluate that the sequence possesses a certain
attribute that a truly random sequence would be likely to exhibit.  
Although various kind of statistical tests are proposed 
so far~\cite{ref:diehard,ref:knuth,ref:fips}, we will focus on NIST
800-22 statistical test suite~\cite{ref:NIST} in this paper 
because this test suite was used for the evaluation of AES candidates.

Some statistical tests are based on a statistical hypothesis $H_0$ which
is that a given binary sequence was produced by a random bit generator.
The test only provides {\it P-value} which is a measure of the strength 
of the evidence provided by the data against the hypothesis.
The significance level $\alpha$ of the test of a statistical hypothesis
$H_0$ is the probability of rejecting $H_0$ when it is true.
If P-value $\geq$ $\alpha$, then the hypothesis $H_0$ is accepted, i.e.,
the sequence would be considered to be random with a confidence $1-\alpha$.
If P-value $<$ $\alpha$, then the hypothesis $H_0$ is rejected, i.e.,
the sequence would be considered to be non-random with a confidence $1-\alpha$.

If the significance level $\alpha$ of a test of $H_0$ is too high, then
the test may reject sequences that were, in fact, produced by a random
bit generator (such an error is called a {\it Type I error}) .
On the other hand, if the significance level $\alpha$ of a test of $H_0$
is too low, then there is the danger that the test may accept sequences
even though they were not produced by a random bit generator (such an
error is called a {\it Type II error}).
It is, therefore, important that the test be carefully designed to have
a significance level that appropriate for the purpose at hand.
However, the calculation of the Type II error is more difficult than the
calculation of $\alpha$ because many possible types of non-randomness
may exists.
Therefore, NIST statistical test suite, which includes 16 tests, adopts 
two further analyses in order to minimize the probability of accepting a 
sequence being produced by a good generator when the generator was 
actually bad~\cite{ref:power}.
First, For each test, a set of sequences (sample size $m$) from output is 
subjected to the test, and the proportion of sequences whose
corresponding P-value satisfies P-value $\geq$ $\alpha$ is calculated.
If the proportion (success rate) is close to $1-\alpha$, then the test
is passed, i.e., the set of sequences is accepted.
Second, the distribution of P-values is calculated for each test.
And, if these P-value are uniformly distributed (no obvious bias), 
then the test is passed.
These two analyses are the crucial difference from the other statistical
test suite.

In section 2, we investigate the randomness of sequences generated by 
various PRBGs including cellular automata (CA)-based PRBG 
using the statistical test suite provided by NIST, and show that 
results of Discrete Fourier Transform (DFT) test and Lempel-Ziv
Compression test are strange.
This suggests that the NIST test setting of these two tests are wrong.
In fact, we identify two mistakes in the NIST setting of DFT test in
section 3.
We also identify two mistakes in the NIST setting of Lempel-Ziv test in
section 4.
The corrections are also given in each section.
This study is important because this NIST test suite was 
used for the evaluation of AES candidates.

\subsection{NIST Statistical Test Suite}

The NIST statistical test suite is a statistical package consisting of 16 tests that
were developed to test the randomness of arbitrary long binary sequences
produced by either hardware or software based cryptographic random or
pseudorandom number generators.
These tests focus on a variety different types of non-randomness that
could exist in a sequence.
The 16 tests are listed in Table 1. 

\begin{table}[ht]
\begin{center}
\caption{List of NIST Statistical Tests}
\vspace{1mm}
\begin{tabular}{c|c} \hline \hline
Number & Test Name \\ \hline \hline
1 & Frequency \\ \hline
2 & Block Frequency \\ \hline
3 & Runs \\ \hline
4 & Longest Run \\ \hline
5 & Binary Matrix Rank \\ \hline
6 & Discrete Fourier Transform \\ \hline
7 & Non-overlapping Template Matching  \\ \hline
8 & Overlapping Template Matching  \\ \hline
9 & Universal  \\ \hline
10 & Lempel Ziv Compression \\ \hline
11 & Linear Complexity  \\ \hline
12 & Serial \\ \hline
13 & Approximate Entropy \\ \hline
14 & Cumulative Sums \\ \hline
15 & Random Excursions \\ \hline
16 & Random Excursions Variant \\ \hline
\end{tabular}
\end{center}
\label{table1}
\end{table}

For each statistical test, a set of P-values, which is corresponding to
the set of sequences, is produced.
Each sequence is called {\it success} if the corresponding P-value
satisfies the condition P-value $\geq$ $\alpha$,
 and is called {\it failure} otherwise.
For a fixed significance level $\alpha$, $100\alpha$ $\%$ of P-values
are expected to indicate failure\footnote{All the statistical tests of
the NIST statistical test suite have the unique significance level $\alpha=0.01$.}.
For the interpretation of test results, 
NIST adopts following two approaches, 
\vspace{1mm}

\noindent
(1) the examination of the proportion of success-sequences (Success Rate)
\vspace{1mm}

If the proportion of success-sequences falls outside of following 
    acceptable interval, there is evidence that the data is non-random.
\begin{equation}
P^{\prime} \pm 3 \sqrt{\frac{P^{\prime} (1 - P^{\prime})}{m}}, 
\label{eq4}
\end{equation}
where $P^{\prime}=1-\alpha$ and $m$ is the number of sequences. 
This interval is determined to be 99.73\% range of normal distribution 
which is an approximation of the binomial distribution under the assumption 
that each sequence is independent sample.
\vspace{2mm}

\noindent
(2) uniformity of the distribution of P-values

This examination is accomplished by computing following $\chi^{2}$ value,
\begin{equation}
\chi^{2} = \sum_{i=1}^{10} \frac{(F_i - m/10)^{2}}{m/10},
\label{eq5}
\end{equation}
where $F_i$ is the number of P-values in sub-interval 
[(i-1)*0.1, i*0.1), and $m$ is the number of sequences (sample size).  
And, the P-value of P-values is calculated such that P$^{\prime}$-value $=$
{\bf igamc} $(9/2, \chi^{2}/2)$, where {\bf igamc}(n,x) is the Incomplete Gamma
Function. 
If P$^{\prime}$-value $\geq$ $0.0001$, then the set of P-values can be
considered to be uniformly distributed.

\section{Results of the NIST Statistical Test Suite}

In this section, we show the results of the NIST statistical 
test suite for four PRBGs (AES, SHA1, MUGI, and CA).
For each statistical test, two further analyses described above are executed, 
and evaluate the set of sequences. 
We use 1000 samples of $10^6$ bit sequences for each test. 
Consequently, 10 (keys) $\times$ 1000
(sample) $\times$ $10^6$ (sequence) bits are used for each test in order
to investigate the difference of the results between different 
keys\footnote{The key is the initial configuration $\{ S^{t=0}_i \} $ in CA case.}.
The input parameters we use are listed in Table 2.
\begin{table}[ht]
\begin{center}
\caption{Parameters used for NIST Test Suite}
\begin{tabular}{c|c} \hline \hline
Test Name & Block Length \\ \hline \hline
Block Frequency & 20,000 \\ \hline
Non-overlapping Template Matching  & 9 \\ \hline
Overlapping Template Matching  & 9 \\ \hline
Universal (Initialization Steps)  & 7 (1280) \\ \hline
Linear Complexity & 500 \\ \hline
Serial & 10 \\ \hline
Approximate Entropy & 10 \\ \hline
\end{tabular}
\end{center}
\label{table2}
\end{table}

\begin{figure}
\centerline{\psfig{figure=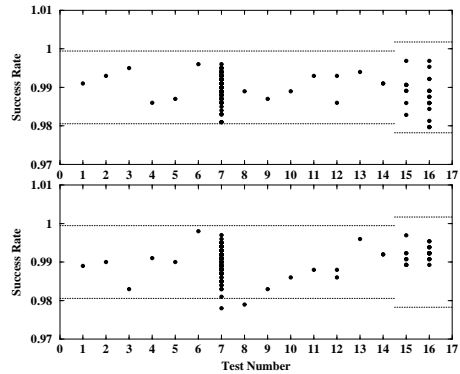,height=5cm}}
\caption{Success rates of AES for 16 tests.
Key 1 (best) and key 7 (worst) cases are shown in up and down figures, respectively.
Dotted lines denote the acceptable interval (eq.(\ref{eq4}) with $\alpha=0.01$).}
\label{fig:AES}
\end{figure}

Table 3 shows the results of AES (128 bit key, OFB mode).
\begin{table}[ht]
\begin{center}
\caption{Results of AES.}
\begin{tabular}{|c|c|c|} \hline \hline
Key & Success Rate  & Uniformity \\ \hline \hline
1 & pass & pass \\ \hline
2 & pass & pass \\ \hline
3 & 15  & pass \\ \hline
4 & pass & pass \\ \hline
5 & 7 & 10 \\ \hline
6 & 14 & 10 \\ \hline
7 & 7, 8 & pass \\ \hline
8 & pass & pass \\ \hline
9 & pass & 10 \\ \hline
10 & pass & 10 \\ \hline
\end{tabular}
\end{center}
\label{table3}
\end{table}
All 16 tests are passed in four cases (key 1, key 2, key 4, and
key 8).
The success rates of the best case (key 1) and of the worst case (key 7)
are shown in Figure \ref{fig:AES}. 
Dotted lines denote the acceptable interval specified by eq.(\ref{eq4}).
As we can see, some tests have many success rates.
For example, the non-overlapping template matching test (number 7) has
148 success rates because one success rate corresponds to the one
template (non-periodic pattern consisting of 9 bits) matching.
If at least one success rates is out of the acceptable interval, then
the test is not passed (see key 7 case).
While all tests are passed in key 1 case, 
the non-overlapping template matching test (number 7) and the overlapping 
template matching test (number 8) are not passed in key 7 case.
It is noted that the uniformity of P-values are not passes only for 
the Lempel-Ziv test (number 10).
The reason why this test is not passed frequently will be explained later.

A one-dimensional 5-neighborhood CA consist of a line
of cells with value $S_i$$=$ 0 or 1 for $i=0,1,2,\cdots,N$.
These cell values are updated in parallel in discrete time steps according to
a fixed rule of the form,
\begin{equation}
S_i^{t+1} = F (S_{i-2}^t, S_{i-1}^t, S_i^t, S_{i+1}^t, S_{i+2}^t),
\label{eq1}
\end{equation}
where $S_i^t$ denotes the $i$ cell value at time 
$t$~\cite{ref:wolfram1,ref:wolfram2,ref:wolfram3}.
We use following rule 535945230 as a CA-based PRBG~\cite{ref:CA}.
\begin{eqnarray}
S_i^{t+1} & = & S_{i-2}^t \oplus S_{i+1}^t \oplus S_{i+2}^t \oplus \nonumber \\ 
 & & S_{i-1}^t \cdot S_{i+1}^t  \oplus S_{i-1}^t \cdot S_{i+2}^t \oplus S_{i}^t \cdot S_{i+1}^t    \oplus \nonumber \\
 & & S_{i}^t \cdot S_{i+2}^t \oplus S_{i+1}^t \cdot S_{i+2}^t \oplus \\
 & & S_{i-1}^t \cdot S_{i+1}^t \cdot S_{i+2}^t      
\oplus S_{i}^t \cdot S_{i+1}^t \cdot S_{i+2}^t \nonumber.
\end{eqnarray}
Table 4, 5 and 6 show the results of SHA1, MUGI, and CA, respectively.
In CA case, we use the cell values $\{ S_i^t \}$ with fixed cell number
$i$ as a sequence, and also use the system size $N=1000$ and 
periodic boundary condition $S_{1}^t=S_{N+1}^t$. 
\begin{table}[ht]
\begin{center}
\caption{Results of SHA1}
\begin{tabular}{|c|c|c|} \hline \hline
Key & Success Rate  & Uniformity \\ \hline \hline
1 & pass & pass \\ \hline
2 & pass & 10 \\ \hline
3 & 7 & pass \\ \hline
4 & 7 & 6 \\ \hline
5 & pass & 10 \\ \hline
6 & 7, 15, 16 & pass \\ \hline
7 & 7 & pass \\ \hline
8 & 7 & pass \\ \hline
9 & pass & pass \\ \hline
10 & pass & 10 \\ \hline
\end{tabular}
\end{center}
\label{table4}
\end{table}
\begin{table}[ht]
\begin{center}
\caption{Results of MUGI}
\begin{tabular}{|c|c|c|} \hline \hline
Key & Success Rate  & Uniformity \\ \hline \hline
1 & 7 & pass \\ \hline
2 & pass & 10 \\ \hline
3 & 10 & 10 \\ \hline
4 & pass & pass \\ \hline
5 & 7 & pass \\ \hline
6 & pass & pass \\ \hline
7 & pass & pass \\ \hline
8 & pass & pass \\ \hline
9 & 7 & pass \\ \hline
10 & pass & 6 \\ \hline
\end{tabular}
\end{center}
\label{table5}
\end{table}
\begin{table}[ht]
\begin{center}
\caption{Results of CA-535945230}
\vspace{1mm}
\begin{tabular}{|c|c|c|} \hline \hline
Key & Success Rate  & Uniformity \\ \hline \hline
1 & pass & pass \\ \hline
2 & pass & 10 \\ \hline
3 & pass & pass \\ \hline
4 & pass & 6, 10 \\ \hline
5 & pass & pass \\ \hline
6 & pass & pass \\ \hline
7 & pass & 7 \\ \hline
8 & pass & pass \\ \hline
9 & pass & 10 \\ \hline
10 & pass & pass \\ \hline
\end{tabular}
\end{center}
\label{table6}
\end{table}
As we can see, all tests are passed in two cases
(SHA1), in four cases (MUGI), and six cases (CA), respectively.
It is noted that results of CA-535945230 case is better than the cases of 
well-known good PRBGs such as AES, SHA1, and MUGI.

\begin{figure}
\centerline{\psfig{figure=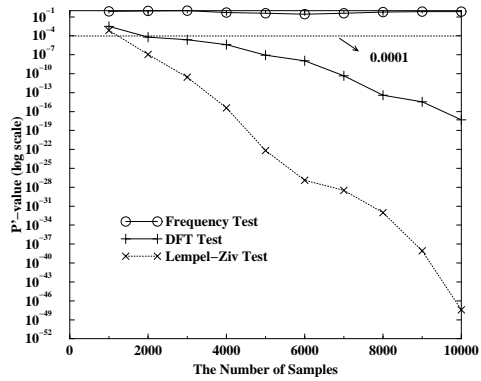,height=5cm}}
\caption{The uniformity of P-values in SHA1 case.}
\label{fig:SHA1}
\end{figure}

If we focus on the uniformity of P-values, only the DFT test (number 6) 
and Lempel-Ziv test (number 10) are not passed frequently.
If we choose the sample size $m$ greater than $10000$, we cannot find 
any PRBGs that pass these two tests even in SHA1 (SHA1 is
used for the mean-value and the variance-value in the distribution of 
the Lempel-Ziv test \cite{ref:NIST}).
Figure \ref{fig:SHA1} shows that P$^{\prime}$-values (the
uniformity of the distribution of P-values) of these two tests 
rapidly decrease as the number of samples increases.
In other words, these distributions of P-values indicate a apparent 
deviation from randomness although we use well-known good PRBG (SHA1).  
This observation suggests that these two tests can be consider as 
an underdeveloped statistical test.
Since many statistical tests are based upon asymptotic approximations,
careful work needs to be done to determine how good an approximation is.
However, we originally found that these two tests have not only 
approximation problem but also mistakes in theoretical setting.

\section{Corrections of Discrete Fourier Transform (Spectral) Test}

In this section, we focus on the DFT test, and show two mistakes
found in the NIST test setting. 
The focus of this test is the peak heights in the Discrete Fourier
Transform of the sequence. 
The purpose of this test is to detect periodic features in the tested
sequence that would indicate a deviation from the assumption of
randomness.
The intention is to detect whether the number of peaks exceeding the
95\% threshold is significantly different than 5\%.
The test description in the NIST document are follows.   
\begin{enumerate}
\item
The zeros and ones of the input sequence ($\epsilon$) are converted to values of -1
     and +1 to create the sequence $X=x_1,x_2,\cdots,x_n$ where $x_i=2\epsilon_i-1$
\item
Apply a Discrete Fourier Transform on X to produce: $S=DFT(X)$.
A sequence of complex variables is produced which represents periodic
     components of the sequence of bits at different frequencies.
\item
Calculate $M=modulus(S^{\prime})\equiv \mid S^{\prime} \mid$, where $S^{\prime}$
     is the substring consisting of the first $n/2$ elements in $S$, and
     the modulus function produces a sequence of peak heights.
\item
Compute $T=\sqrt{3n}=$ the 95\% peak height threshold value.
Under the assumption of randomness, 95\% of the values obtained from the
     test should not exceed $T$. 
\item
Compute $N_0=0.95n/2$.
$N_0$ is the expected theoretical (95\%) number of peaks that are less
     than $T$.
\item
Compute $N_1=$ the actual observed number of peaks in $M$ that are less
     than $T$.  
\item
Compute $d=\frac{N_1-N_0}{\sqrt{n(0.95)(0.05)/2}}$.
\item
Compute P-value $=erfc(\frac{\mid d \mid}{\sqrt{2}})$.
\end{enumerate}

\subsection{The derivation of the threshold $T$}

First, we show the derivation of the threshold $T=\sqrt{3n}$.
For a frequency $j$, DFT are defined by following equation.
\begin{equation}
S_j = \sum_{k=1}^{n} x_k cos ( 2\pi\frac{(k-1)}{n}j ) + i \sum_{k=1}^{n} x_k sin(2\pi\frac{(k-1)}{n}j) .
\end{equation}
Let us consider the square of modulus of $S_j$, 
\begin{equation}
{\mid S_j \mid}^2 = c_j^2 + s_j^2
\end{equation}
, where 
\begin{eqnarray}
c_j & = & \sum_{k=1}^n x_k cos ( 2\pi\frac{(k-1)}{n}j )\\
s_j & = & \sum_{k=1}^n x_k sin ( 2\pi\frac{(k-1)}{n}j ) .
\end{eqnarray}
Here, we can simply prove that $c_j$ and $s_j$ converge to the normal
distribution whose mean $\mu$ is zero and variance $\sigma^2$ is $n/2$ under the
assumption of $x_k$ ($-1$ or $+1$ for $k=1,2,\cdots,n$) randomness.
Therefore, $Y$ $=$ $(\frac{c_j}{\sigma})^2$ $+$ $(\frac{s_j}{\sigma})^2$
converges to following distribution function ($\chi^2$ distribution with
2 degree of freedom),
\begin{equation}
P(Y) = \frac{1}{2} \exp(-\frac{Y}{2}) .
\end{equation}
If we transform $Y$ to $Z=\frac{Y}{2}$, we can get following
distribution,
\begin{equation}
P(Z) = \exp(-Z) .
\end{equation}
The threshold $T$ is defined such that the number of peaks exceeding the
threshold $T$ should be 5\% under the assumption of randomness.
Since 
\begin{equation}
\int_{Z_C}^{\infty} \exp (-Z) dZ = \exp (-Z_C) = 0.05,
\end{equation}
we can get the value $Z_C = -ln (0.05) = 2.995732274$.
From $\mid S_j \mid$ $=$ $\sqrt{nZ}$, we conclude that
\begin{equation}
T = \sqrt{2.995732274 n} .
\end{equation}

\begin{figure}
\centerline{\psfig{figure=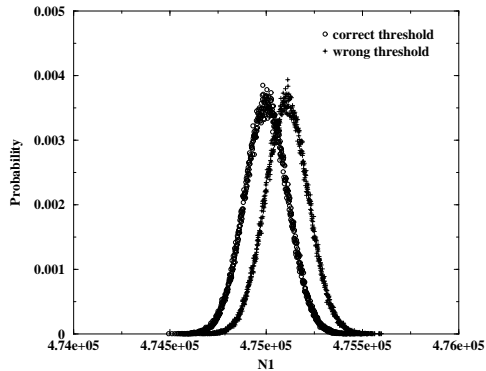,height=5cm}}
\caption{The distribution of $N_1$ in SHA1
case ($300000$ samples of $n=10^6$ bit sequence).
Note that the expected value of $N_1$, that is, $N_0$ is $475000$.}
\label{fig:N1}
\end{figure}

We have found that the deviation of $\sqrt{3n}$ from 
$\sqrt{2.995732274n}$ makes the distribution invalid.
Figure \ref{fig:N1} shows the distribution of $N_1$ in SHA1
case ($300000$ samples of $n=10^6$ bit sequence).
Note that the expected value of $N_1$, that is, $N_0=\frac{0.95n}{2}$ 
is $475000$ in this case. 
If we set the threshold $T=\sqrt{3n}$, then the distribution is shifted
to the right.
So, we have to set the threshold $ T = \sqrt{2.995732274 n} $.
This is the first mistakes in DFT test.

\subsection{The theoretical distribution}

\begin{figure}
\centerline{\psfig{figure=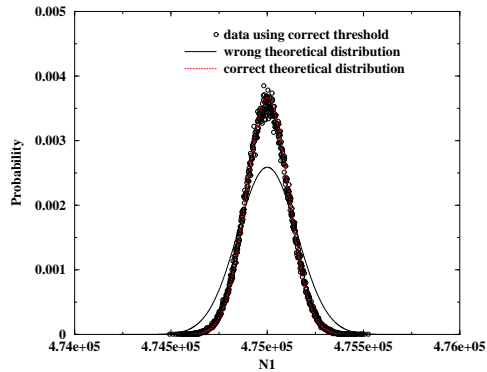,height=5cm}}
\caption{The fitting of the distribution of $N_1$ in SHA1 case 
with the threshold $ T = \sqrt{2.995732274n} $ and two theoretical 
distributions.}
\label{fig:N1fitting}
\end{figure}

Because we use the real values $x_k$, the symmetry such as 
$\mid S_j \mid$  $=$ $\mid S_{n-j} \mid$ appears in peaks.
So, the NIST focus on the first $\frac{n}{2}$ peaks. 
The test description in the NIST documents use the theoretical 
distribution whose mean value $\mu$ is $\frac{np}{2}$ and variance value
$\sigma^2$ is $\frac{npq}{2}$ where $p=0.95$, $q=0.05$, and $n=10^6$
($\frac{n}{2}$ times coin tossing with probability $p$ and $q$).
However, this coin tossing is not independent process.
The quantity $\sum_{j}^{n/2} S_j$ is conserved in this process.
In this case, the variance $\sigma^2$ becomes $\frac{npq}{4}$.
Figure \ref{fig:N1fitting} shows the fitting of the distribution of 
$N_1$ in SHA1 case with the threshold 
$ T = \sqrt{2.995732274n} $ and two theoretical distributions.
We can confirm that the distribution becomes to fit to the new 
theoretical distribution.

\section{Corrections of Lempel-Ziv Compression Test}

In this section, we focus on the Lempel-Ziv test, and show two mistakes
found in the NIST test setting. 
The focus of this test is the number of cumulatively distinct patterns
(words) in the sequence.
The purpose of the test is to determine how far the tested sequence can
be compressed.
The sequence is considered to be non-random if it can be significantly
compressed.
A random sequence will have a characteristic number of distinct
patterns.
The test description in the NIST document are follows.   
\begin{enumerate}
\item
Parse the sequence into consecutive, disjoint and distinct words that
     will form a ``dictionary'' of words in the sequence.
This is accomplished by creating substrings from consecutive bits of the
     sequence until a substring is created that has not been found
     previously in the sequence.
The resulting substring is a new word in the dictionary. 
\item
Compute P-value 
$= \frac{1}{2} erfc(\frac{\mu-W_{obs}}{\sqrt{2\sigma^2}})$,\\
 where $\mu=69588.2019$ and $\sigma^2=73.23726011$ when $n=10^6$
(these values are updated Oct. 26, 1999).
Note that since no known theory is available to determine the exact
     values of $\mu$ and $\sigma$, these values were computed using
     SHA1.   
\end{enumerate}

\subsection{The asymmetric distribution}

\begin{figure}
\centerline{\psfig{figure=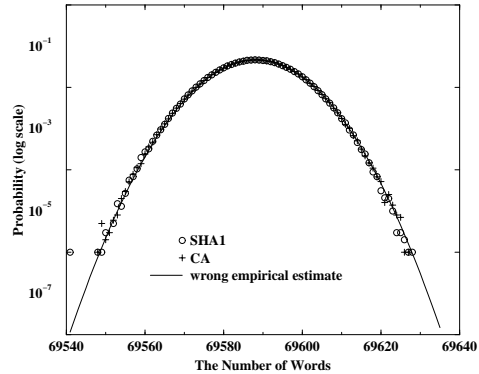,height=5cm}}
\caption{The distribution of the number of words in SHA1 case 
and CA case.
$10^6$ samples of $n=10^6$ bit sequence are used.}
\label{fig:LZ}
\end{figure}

There are asymptotically well-approximated mean formula and the variance
formula of the distribution of the Lempel-Ziv test \cite{ref:LZ1,ref:LZ2}.
However, it is known that above formulas are invalid for the sequence of 
length less than $10^7$ through a simulation study using BBS. 
Therefore, SHA1, which is one of well-known good PRBGs, is used instead for 
the mean-value and the variance-value in the NIST setting \cite{ref:NIST}. 
The accuracy of such empirical estimates depends on the randomness of
the generator used.
Figure \ref{fig:LZ} shows the distributions of the number of words in SHA1 case 
and CA case ($10^6$ samples of $n=10^6$ bit sequence).
Two distributions are almost the same although two algorithms are
completely different.
\begin{figure}
\centerline{\psfig{figure=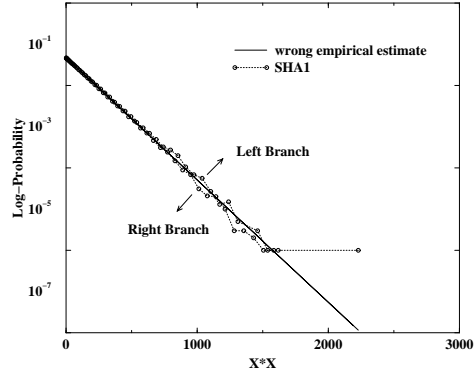,height=5cm}}
\caption{The distribution of the number of words (SHA1 case) in
 different scale. The horizontal axis denotes the square of distance from
 the mean value for both branchs. The same data of Fig. \protect{\ref{fig:LZ}} is used.}
\label{fig:X2}
\end{figure} 
We can confirm the subtle asymmetries if we see Fig. \ref{fig:X2} carefully.
We conclude that this distribution can be used for the mean and
variance values of new setting of the test.
Through the fitting of the distributions, we got the mean value $\mu$
$=$ $69588.09$ and variance values $\sigma_{L}^2$ $=$ $75.574336518$ 
and $\sigma_{R}^2$ $=$ $72.42178447$, for the left branch and right
branch, respectively.
Consequently, we got the new empirical estimates (asymmetric
distribution) which are better than the NIST setting.

\subsection{The effect of discreteness}

\begin{figure}
\centerline{\psfig{figure=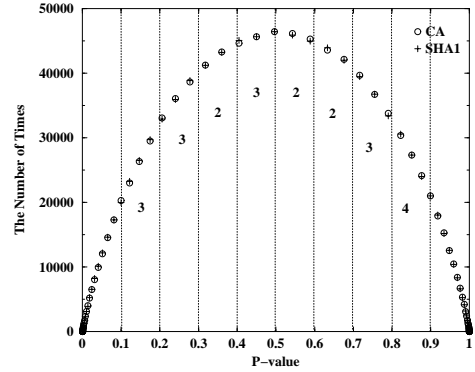,height=5cm}}
\caption{The number of times of appeared P-values in SHA1 and CA cases.
$10^6$ samples of $n=10^6$ bit sequence are used. 
The numbers described in figure denote the variety of appeared P-values in each bin.}
\label{fig:discrete}
\end{figure}

Despite the best fitting of the distribution, the uniformity of P-values
can not be improved.
This is because the distribution of the number of words is too narrow 
(the variance is too small).
Therefore, the effect of discreteness appeared.
In other words, a variety of the appeared P-values is limited.
Figure \ref{fig:discrete} shows the number of times of appeared P-values
in SHA1 and CA cases.
Because the variety of appeared P-values are two or three in centered bins, 
we never get the uniformity of P-values in this situation.

Because the purpose of checking the uniformity of P-value is to detect the
deviation of the distribution from that of random sequence case, we
re-define the uniformity of P-values only in this test case as the 
histogram of P-values itself which is produced by SHA1 and 
CA5 ($10^6$samples).
In other words, we use following formula for the checking of the
uniformity instead of eq.(\ref{eq5}),
\begin{equation}
\chi^{2} = \sum_{i=1}^{10} \frac{(F_i - m S_i)^{2}}{m S_i},
\end{equation}
where $m$ denotes sample size and $S_i$ denotes the rate of each $i$ bin
which is computed from the histogram of P-values ($10^6$ samples of SHA1
and CA data), that is,  $S_1=0.1097085$, $S_2=0.079127$, $S_3=0.107691$,
$S_4=0.084465$, $S_5=0.1369235$, $S_6=0.091115$, $S_7=0.0858035$,
$S_8=0.1098615$, $S_9=0.1028565$, and $S_{10}=0.0924485$.

\section{Conclusion}

\begin{figure}
\centerline{\psfig{figure=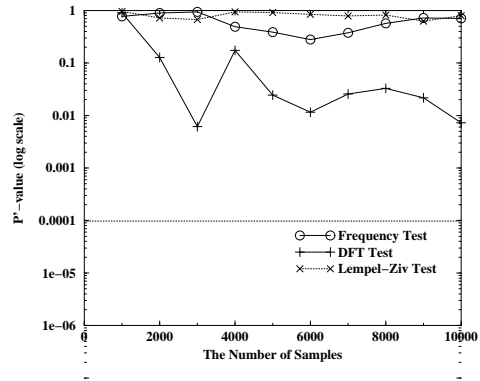,height=5cm}}
\caption{The improved uniformity of P-values in SHA1 case.}
\label{fig:con}
\end{figure}

We corrected two points for DFT test setting,
\begin{enumerate}
\item
The correction of the threshold $T$ from $\sqrt{3n}$ to $\sqrt{2.995732274 n}$.
\item
The correction of the variance $\sigma^2$ of theoretical distribution
     from $\frac{npq}{2}$ to $\frac{npq}{4}$. 
\end{enumerate}
We also corrected two points for Lempel-Ziv test,
\begin{enumerate}
\item
The setting of standard distribution which has no algorithm dependence.
This asymmetric normal distribution has its mean value $\mu$ $=$
     $69588.09$ and variance values 
$\sigma_{L}^2$ $=$ $75.574336518$ and $\sigma_{R}^2$ $=$ $72.42178447$, 
for the left branch and right branch, respectively, in $n=10^6$ case.
\item
the re-definition of the uniformity of P-values as the histogram of
     P-values itself which is produced by SHA1 and CA5 ($10^6$samples).
\end{enumerate}
Figure \ref{fig:con} shows the P$^{\prime}$-values behavior 
after corrections when the number of samples increases.
As a result, P$^{\prime}$-values of two test become improved 
(compare with Fig. \ref{fig:SHA1}).  

Although the checking of the uniformity of P-values was not executed in
the evaluation of AES candidate algorithms, the used P-value itself has
nonsense in these two tests.
This suggests that re-evaluation of the test results should be needed.

%

\end{document}